\documentclass[aps,prl,twocolumn,showpacs]{revtex4}
\usepackage{graphicx}
\begin{document}
\title{Ortho and Para Molecules of Water in Electric Field}
\author{S. N. Andreev, V. P. Makarov, V. I. Tikhonov, and A. A. Volkov}
\address{A. M. Prokhorov General Physics Institute, Russian Academy of Sciences,
Vavilov Street 38, 119991 Moscow, Russia }

\date{\today}

\begin{abstract}

Stark effect is calculated by the perturbation theory method
separately for the \emph{ortho} and \emph{para} water molecules.
At room temperature, a 30\%-difference in the energy change is
found for the two species put in electric field. This implies a
sorting of the \emph{ortho} and \emph{para} water molecules in
non-uniform electric fields. The \emph{ortho}/\emph{para} water
separation is suggested to occur in the course of steam sorption
on a solid surface and of large-scale atmospheric processes.

\end{abstract}


\maketitle

It is a textbook knowledge that an isolated water molecule,
H$_{2}$O, exists in one of the two forms (spin isomers),
\emph{ortho} or \emph{para}, with parallel or antiparallel proton
spins, respectively \cite{townes}. In the first case a total
magnetic moment of the molecule is 1, while in the second case it
is 0. Under ambient conditions the gaseous \emph{ortho} and
\emph{para} molecules are in a statistical 3:1 equilibrium
(further referred as the normal O/P-ratio). In the gas, the
\emph{ortho}-\emph{para} conversion due to collisions and
radiation processes is highly improbable \cite{oka}. The valuable
factor of conversion is magnetic interactions. In the absence of
magnetic impurities, the estimated life-time of the gaseous
\emph{ortho} and \emph{para} molecules is months \cite{curl}.
Thus, the water vapor is a true mixture of two gaseous ensembles
consisting of magnetic (\emph{ortho}) and nonmagnetic
(\emph{para}) molecules. This means that the \emph{ortho} and
\emph{para} ensembles can in principle be spatially separated.

For molecular hydrogen, the task of spatial separation of the spin
isomers has been solved decades ago \cite{farkas}. At present, the
O/P separation of H$_{2}$ is usually attained by catalytically
assisted cooling of gaseous H$_{2}$ at cryogenic temperatures.
Separate existence of \emph{ortho} and \emph{para} water, however,
is still questionable. The challenge is to pinpoint those physical
properties of the chemically identical \emph{ortho} and
\emph{para} water molecules, which can be used for their
separation. Here, by the perturbation theory calculations we study
response of the \emph{ortho} and \emph{para} water molecules to
the dc electric field (Stark effect), and find that these two
species have noticeably different electrostatic properties.
Although the study of Stark effect in water has a long history
\cite{townes, mengel}, the effect, to our knowledge, has never
been considered separately for the \emph{ortho} and \emph{para}
modifications.

As is generally accepted, we consider a water molecule as a rigid
asymmetrical top and choose the coordinate system so that the
$\xi, \eta, \zeta$ axes are directed along the three main inertial
axes of the molecule \cite{townes}. Further we follow a standard
procedure described in details in Ref. \cite{landau}. The
rotational part of Hamiltonian is written as:

\begin{equation}
\hat{H}_{rot}=\frac{\hbar^{2}}{2}\cdot(\frac{\hat{J_{\xi}^{2}}}{I_{A}}+\frac{\hat{J_{\eta}^{2}}}{I_{B}}+\frac{\hat{J_{\zeta}^{2}}}{I_{C}}),
\label{hamiltonian}
\end{equation}
where  $\hat{J_{\xi}},\hat{J_{\eta}},\hat{J_{\zeta}}$ are the
angular momentum operators, and $I_{A}, I_{B}, I_{C}$ are the main
inertial moments of the molecule. The symmetry of the Hamiltonian
(\ref{hamiltonian}) belongs to the $D_{2}$ group, thus the
rotational energy levels are classified according to the
irreducible representations of this group: $A$, $B_{1}$, $B_{2}$
and $B_{3}$. After rotation by $\pi$ along the $\zeta$ direction
($C_{\pi}(\zeta)$), the wave functions of the $A$ and $B_{1}$
symmetries do not change their signs, while the wave functions of
the $B_{2}$ and $B_{3}$ symmetries do.

In the Born-Oppenheimer approximation and neglecting the
rotation-vibration interaction, the wave function of water
molecule is
$|\Psi\rangle=|\psi_{s}\rangle\cdot|\psi_{el}\rangle\cdot|\psi_{vibr}\rangle\cdot|\psi_{rot}\rangle$,
where $|\psi_{s}\rangle,|\psi_{el}\rangle,|\psi_{vibr}\rangle$,
and $|\psi_{rot}\rangle$  are the nuclear spin, electronic,
vibrational, and rotational wave functions. According to the Pauli
principle, the total $\Psi$-function should change its sign after
permutation of the hydrogen atoms. This permutation is equivalent
to the $C_{\pi}(\zeta)$ rotation. The functions
$|\psi_{el}\rangle$ and $|\psi_{vibr}\rangle$ of the ground state
are symmetric with respect to $C_{\pi}(\zeta)$. Whether or not
$|\psi_{rot}\rangle$ and $|\psi_{s}\rangle$ change their signs,
depends on in which states, \emph{ortho} or \emph{para}, the water
molecule is. Obviously, for the case of the parallel proton spins
(the \emph{ortho} molecule) the exchange of the protons does not
reverse $|\psi_{s}\rangle$. In this case, only
$|\psi_{rot}\rangle$ is responsible for the sign inversion of the
total $\Psi$-function. Therefore, $|\psi_{rot}\rangle$ of the
\emph{ortho} molecule has either $B_{2}$ or $B_{3}$ symmetries. In
contrast, $|\psi_{s}\rangle$ of the \emph{para} molecule does
change the sign when the protons (with antiparallel spins) are
interchanged. In turn, $|\psi_{rot}\rangle$ remains sign-constant.
Hence, the \emph{para} rotational states have wave functions of
the $A$ and $B_{1}$ types.

We calculate the rotational levels and corresponding wave
functions $|\psi_{rot}\rangle$ using the perturbation theory
method. The electric field is assumed to be weak enough so that
the level shifts caused by the field are much smaller than the
nominal distances between the levels. The direction of the
$z$-axis of the immovable laboratory coordinate system $(x,y,z)$
is chosen along the electric field vector. The interaction energy
of the water molecule with electric field is:

\begin{equation}
V=-\mathcal{E} \cdot d_{z}, \label{interaction_energy}
\end{equation}
where $d_{z}$ is the projection of the dipole moment of the water
molecule to the $z$-axis. In the first approximation, the shifts
of the levels are determined by the diagonal matrix elements of
the operator (\ref{interaction_energy}). However, they are equal
to zero for the asymmetrical rotators. Thus, the level splitting
of the water molecule is a second order effect relative to the
electric field:

\begin{equation}
\Delta E_{JM_{J}n}= \mathcal{E}^{2} \sum_{J'M_{J'}n'} \frac
{|D|^{2}}{E_{Jn}-E_{J'n'}}, \label{delta_E}
\end{equation}
where $D=\langle J'M_{J'}n'|d_{z}|JM_{J}n\rangle$ is a matrix
element of $d_{z}$, $J$ is a total moment of the molecule, $M_{J}$
is its projection to the z-axis, the index $n$ numerates different
states with given $J$ and $M_{J}$. The rotational wave function of
the asymmetrical rotator is  $|JM_{J}n\rangle=\sum_{k}
C^{(Jn)}_{k}|JM_{J}k\rangle$, where $|JM_{J}k\rangle$ are
eigenfunctions of the operators $\hat{J}^{2}$, $\hat{J}_{z}$, and
$\hat{J}_{\xi}$, with eigenvalues, correspondingly, $J(J+1)$,
$M_{J}=J,J-1,...,-J$, and $k=J,J-1,...,-J$. Then $D$ can be
expressed in terms of the eigen dipole moment of the molecule,
$d^{(0)}$, directed along the $\zeta$ axis, in a following way:

\begin{eqnarray}
D=\sum_{kk'}C_{k'}^{(J'n')*}C_{k}^{(Jn)}\langle
J'M_{J'}k'|d_{z}|JM_{J}k\rangle=\nonumber\\
d^{(0)}\cdot\sum_{kk'}C_{k'}^{(J'n')*}C_{k}^{(Jn)}(i)^{J-J'}(-1)^{k-M_{J'}}\times\nonumber\\
\sqrt{(2J+1)(2J'+1)}\cdot\left( \begin{array}{ccc} J' & J & 1 \\
-k & k & 0 \end{array} \right) \times \nonumber\\
\left( \begin{array}{ccc} J' & J & 1 \\ -M_{J} & M_{J} & 0
\end{array} \right) \cdot \delta_{kk'}\delta_{M_{J}M_{J'}},
\label{dipole}
\end{eqnarray}
where $\left( \begin{array}{ccc} J' & J & 1 \\
-k & k & 0 \end{array} \right)$ and $\left( \begin{array}{ccc} J'
& J & 1 \\ -M_{J} & M_{J} & 0
\end{array} \right)$ are the Klebsh-Gordon
coefficients which are non-zero only for $J'=J$ and $J'=J\pm1$. In
view of expression (\ref{dipole}), the general formula
(\ref{delta_E}) is reduced to:

\begin{equation}
\Delta E_{JM_{J}n}= -\frac{1}{2} \mathcal{E}^{2} \left(
\alpha_{Jn}+2\beta_{Jn}\left(
M_{J}^{2}-\frac{1}{3}J(J+1)\right)\right), \label{delta_E1}
\end{equation}
where $\alpha_{Jn}$ is the total shift of the "center of gravity"
of the split level and $\beta_{Jn}$ is the shift of the sub-levels
relative to this "center of gravity". From Eqs. (\ref{delta_E})
and (\ref{delta_E1}), by performing summation over $M_{J}$ and
$M_{J'}$, we obtain the expressions for $\alpha_{Jn}$ and
$\beta_{Jn}$. The Stark behavior of the rotational levels of the
water molecule is then calculated with the following parameters
\cite{loesch,tikhonov}: $a=\hbar^{2}/I_{A}=55.7$ cm$^{-1}$,
$b=\hbar^{2}/I_{B}=18.6$ cm$^{-1}$, $c=\hbar^{2}/I_{C}=29.0$
cm$^{-1}$, and $d^{(0)}$ = 1.84 Debye.

\begin{figure}[]
\centering
\includegraphics[width=\columnwidth,clip]{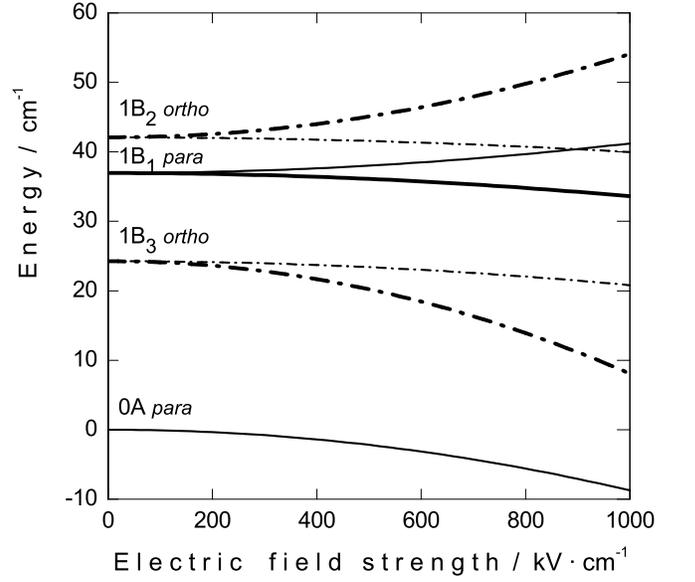}
%\vspace{0.2cm}
\caption{Stark effect for the water molecule: positions of the
lowest four rotational energy levels, $E_{J}$, in external
electric field. Light and bold lines show non-degenerated and
twice degenerated levels, correspondingly.} \label{Stark_effect}
\end{figure}

The results of the calculations for $J=0;1$ are shown in Fig.
\ref{Stark_effect}. The calculations are limited by the field of
$10^{6}$ V/cm to meet the requirements of the second order
approximation method we used. It is seen from the Figure, that the
singlet \emph{para}-level $|0A\rangle$ with $J=0$, shifts down
when the field is applied. This is, of course, in accordance with
the general quantum mechanical theorem \cite{landau}. The levels
with $J=1$ split into 2 sub-levels. One of these two sub-levels
has $M_{J}=0$ and is not degenerated (thin lines in Fig.
\ref{Stark_effect}). Another sub-levels correspond to $M_{J}=\pm1$
and are twice degenerated (bold lines).

Fig. \ref{Stark_effect} reveals qualitatively different trends for
the \emph{ortho} and \emph{para} levels in the field. When the
field increases, the majority of the para-levels (the singlet
$|0A\rangle$ and the twice-degenerated sub-level $|1B_{1}\rangle$)
goes down. Only one of the \emph{para} levels goes up. Thus, the
net energy shift of the \emph{para} levels is appreciably
negative. In the case of \emph{ortho} molecules, the
twice-degenerated \emph{ortho} levels (bold dash-dotted lines)
diverge symmetrically as field goes up, the non-degenerated
\emph{ortho} levels lowering slightly only. As a result, the net
energy shift of the \emph{ortho} levels is notably, approximately
two times, smaller than the net shift of the \emph{para} levels.

Taking into account the Boltzmann distribution of water molecules
over the energy levels, one can write the net energy shifts
$\overline{\Delta E_{O}}$ and $\overline{\Delta E_{P}}$ of the
gaseous \emph{ortho} and \emph{para} molecules in the electric
field $\mathcal{E}$:

\begin{equation}
\overline{\Delta E_{O,P}} = -\frac {1}{2} \mathcal{E}^{2} \frac
{\sum \alpha_{Jn}(2J+1)exp(-\Delta E_{JM_{J}n}/T)}{\sum (2J+1)
exp(-\Delta E_{JM_{J}n}/T)}, \label{delta_E_OP}
\end{equation}
where $\Delta E_{JM_{J}n}=E_{JM_{J}n}-E_{0}$ are the energies of
the \emph{ortho} and \emph{para} levels $E_{J}$ about the energy
$E_{0}$ of the lowest $|1B_{3}\rangle$ and $|0A\rangle$ levels,
correspondingly; $T$ is the temperature; the summation is carried
out separately over the \emph{ortho} and \emph{para} levels.

The ratio $R=\overline{\Delta E_{P}}/\overline{\Delta E_{O}}$
gives a comparative response of the \emph{ortho} and \emph{para}
water molecules to an external electric field. $R$ can be
considered as a ratio of the forces acting on the \emph{para} and
\emph{ortho} water molecules in a non-uniform electric field. As
is seen from Eq. \ref{delta_E_OP}, it is a function of temperature
$T$ and does not depend on the field strength.

\begin{figure}[]
\centering
\includegraphics[width=\columnwidth,clip]{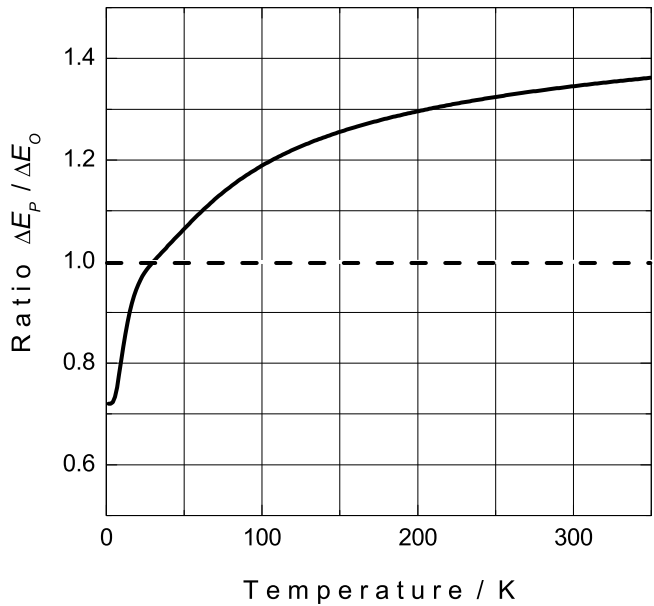}
%\vspace{0.2cm}
\caption{Ratio of the energy Stark shifts for the \emph{para} and
\emph{ortho} water molecules as a function of temperature. The
dashed line notices the unity for eye.} \label{ratio}
\end{figure}

We have calculated $R(T)$ for 16 lowest rotational levels
($J\leq3$), laying in the range of 0-283 cm$^{-1}$ and containing
a 70 \% deal of the molecules at room temperature. The Stark
shifts of the higher levels of the asymmetric top decrease quickly
when $J$ increases \cite{townes}, therefore the contribution of
the higher energy levels in $R(T)$ is comparatively small (less of
5\% at room temperature). The result of our calculations is
presented in Fig. \ref{ratio}.

The drop of the $R(T)$ at $T < 30$ K is easy to understand from
simple considerations. Indeed, at low temperatures the water
molecules occupy mainly the two lowest \emph{ortho}
$|1B_{3}\rangle$ and \emph{para} $|0A\rangle$ levels. In
accordance with Fig. \ref{Stark_effect}, the \emph{ortho}
$|1B_{3}\rangle$ level is of a superior "Stark flexibility". In an
electric field it goes sharply down predominantly causing the
greater energy lowering of the \emph{ortho} molecules.

At elevated temperatures the additional \emph{para} and
\emph{ortho} levels get filled with molecules resulting in
levelling off of the total energy Stark shifts of the \emph{ortho}
and the \emph{para} molecules. Surprisingly, however, the $R(T)$
ratio does not go to the unity, but exceeds it, reaching the 1.34
value at room temperature. This implies that the \emph{para}
molecules in non-uniform electric fields experience a one third
stronger electric force than the \emph{ortho} molecules. The
\emph{para} molecules are thus proved to be more active as the
"high field seeking" molecules.

One can suggest the occurrence of the O/P sorting effect for the
\emph{ortho} and \emph{para} water molecules in non-uniform
electric fields, as it is typically happening with polar molecules
\cite{loesch}. Let us note, it is just electric rather than
magnetic field that leads to a sizable distinction between the
molecules with different magnetic arrangement.

Obviously, one of the fields where the distinction between
electrostatic properties of the \emph{ortho} and \emph{para} water
molecules can naturally be revealed is a water vapor sorption on a
solid surface. The impinging water molecules are unavoidably
exposed to the non-uniform surface electric field and, therefore,
are suggested to undergo the O/P separation effect. A rough
estimate \cite{townes} of the total rotational energy drop
$\Delta$ of the water molecule on the surface is: $\Delta \sim
-d\cdot\mathcal{E} \sim -1$ kJ/mol, where $d$ = 1.84 Debye is a
dipole moment of water molecule and $\mathcal{E} \sim 10^{6}$ V/cm
is taken as an estimate of both, a typical surface electric field
and the highest strength allowed by our perturbation method. As is
seen, the dipole attraction energy is not negligible in comparison
with the real physadsorption energies of 5 - 20 kJ/mol
\cite{linsen}. The 30\% difference in $\Delta$ for the
\emph{ortho} and \emph{para} molecules looks large enough to be
detectable during the adsorption. We suggest that this effect is
responsible for the O/P water separation observed in absorption
experiments with water vapor \cite{tikhonov}.

Beyond the sorption, a sizable effect of the O/P water separation
can be suggested to occur in the atmosphere. In fact, the water
vapor and atmospheric electricity (ranging from 100 V/m to 1000
V/m normally and to millions V/m in lightnings) are inherent
attributes of the atmosphere. Allowance of O/P water separation
may give a new insight to the atmospheric phenomena.

Summarizing, our calculations point to a remarkable difference in
the forces acting on the \emph{ortho} and \emph{para} water
molecules subjected to non-uniform electric field. This
distinction is of a fundamental origin. Therefore, we suggest it
should play an important role in nature environments.

This work is supported by the European Commission (NEST-Adventure
program, grant 5032) and by the Russian Foundation for Basic
Research (grant 06-08-00937a).


\begin{thebibliography}{99}

\bibitem{townes} C. H. Townes and A. L. Schawlow, \textit{Microwave spectroscopy}
(McGraw-Hill Publishing Company, New York - London - Toronto,
1955).

\bibitem{oka} T.Oka, Adv. Atom. and Mol. Phys. \textbf{9}, 127
(1973).

\bibitem{curl} R. F. Curl, J. V. V. Kasper, and K. S. Pitzer, J. Chem. Phys. \textbf{46}, 3220 (1967).

\bibitem{farkas} A. Farkas, \textit{Orthohydrogen, parahydrogen and heavy hydrogen}
(Cambridge, 1935), p. 230

\bibitem{mengel} M. Mengel, P. Jensen, J. Molecular Spectroscopy \textbf{169},
73 (1995).

\bibitem{landau} L. D. Landau and E. M. Lifshitz, \textit{Quantum Mechanics, Nonrelativistic
theory} (Pergamon press, Oxford, 1976).

\bibitem{loesch}  H. J. Loesch and B. Scheel, Phys. Rev. Lett. \textbf{85}, 2709 (2000).

\bibitem{linsen} \textit{Physical and chemical aspects of adsorbents and catalysts},
edited by B. G. Linsen (Academic Press, London - New York, 1970).

\bibitem{tikhonov}  V. I. Tikhonov and A. A. Volkov, Science \textbf{296}, 2363 (2002);
ChemPhysChem \textbf{7}, 1026 (2006).

\end{thebibliography}
\end{document}